\documentclass[12pt,titlepage,letterpaper]{article}

\usepackage{amsmath}

\usepackage[pdftex]{graphicx}

\usepackage[round,numbers,sort&compress]{natbib} 
\usepackage{times}

\usepackage{color}
\newcommand{\colorA}{}
\newcommand{\colorB}{}

\usepackage{hhline}
\usepackage{multirow}

\newcommand{\mr}[1] {\multirow{2}{*}{#1}}
\newlength{\delim}
\newcommand{\delimwidth}{\settowidth{\delim}{$\Big\}$}}
\newcommand{\mrd}[1]{\multirow{2}{*}{\hspace{-\delim}$\Big\}$#1}}

\newcommand{\beq}{\begin{equation}}
\newcommand{\eeq}{\end{equation}}
\newcommand{\bea}{\begin{eqnarray}}
\newcommand{\eea}{\end{eqnarray}}

\newcommand{\kB}{k_\mathrm{B}}
\newcommand{\temp}{T}
\newcommand{\kT}{k_\mathrm{B}\temp}
\newcommand{\e}{\mathrm{e}}

\newcommand{\rec}{{}$^{-1}${}}
\newcommand{\Lp}{l_\mathrm{p}}
\newcommand{\LK}{l_\mathrm{K}}
\newcommand{\Ld}{L_\mathrm{d}}
\newcommand{\Nu}{N_\mathrm{u}}
\newcommand{\Nd}{N_\mathrm{d}}

\newcommand{\vR}{\vec{R}}
\newcommand{\vF}{\vec{F}}
\newcommand{\vL}{\vec{L}}
\newcommand{\vLd}{\vec{L}_\mathrm{d}}

\newcommand{\tr}{\rightleftharpoons}

\newcommand{\dd}{\mathrm{d}}
\newcommand{\sT}{\mathrm{T}}
\newcommand{\sTs}{\mathrm{T^*}}
\newcommand{\sD}{\mathrm{D}}
\newcommand{\sDs}{\mathrm{D^*}}

\newcommand{\sDt}{\mathrm{\widetilde{D}}}
\newcommand{\sE}{\mathrm{0}}
\newcommand{\sTTs}{\sT^\mathrm{(*)}}
\newcommand{\sDDs}{\sD^\mathrm{(*)}}
\newcommand{\sDDst}{\sDDs_\mathrm{t}}
\newcommand{\sDDsl}{\sDDs_\mathrm{l}}
\newcommand{\eg}{e.g.}
\newcommand{\ie}{i.e.}
\newcommand{\etc}{etc}

 \newcommand{\FIGscheme}	{1}
 \newcommand{\FIGtwod}		{2}
 \newcommand{\FIGcarter}	{3}
 \newcommand{\FIGATPvelrand}	{4}
 \newcommand{\FIGmichio}	{5}

 \newcommand{\FIGG}		{1}
 \newcommand{\FIGgtgd}		{2}

 \newcommand{\TABparameters}	{1}

 \newcommand{\TABcriteria}	{1}
 \newcommand{\TABdeviations}	{2}

\title{Neck linker docking coordinates the kinetics of kinesin's heads}

\author{
 Andr\'as Cz\"ovek\\
  {\small Department of Biological Physics}\\
  {\small E\"otv\"os University, Budapest, Hungary}
 \and
 Gergely J. Sz\"oll\H osi\\
  {\small CNRS UMR 5558}\\
  {\small Universit\'e Claude Bernard, Villeurbanne, France}
 \and
 Imre Der\'enyi\thanks{Corresponding author}\\
  {\small Department of Biological Physics}\\
  {\small E\"otv\"os University, Budapest, Hungary}
}

\begin{document}

\maketitle

\abstract{%
Conventional kinesin is a two-headed homodimeric motor protein, which
is able to walk along microtubules processively by hydrolyzing ATP. Its
neck linkers, which connect the two motor domains and can undergo a
docking/undocking transition, are widely believed to play the key role
in the coordination of the chemical cycles of the two motor domains
and, consequently, in force production and directional stepping.
Although many experiments, often complemented with partial kinetic
modeling of specific pathways, support this idea, the ultimate test of
the viability of this hypothesis requires the construction of a
complete kinetic model. Considering the two neck linkers as entropic
springs that are allowed to dock to their head domains and
incorporating only the few most relevant kinetic and structural
properties of the individual heads, here we develop the first detailed,
thermodynamically consistent model of kinesin that can (i) explain the
cooperation of the heads \colorB{(including their gating mechanisms)}
during walking and (ii) reproduce much of the
available experimental data (speed, dwell time distribution,
randomness, processivity, hydrolysis rate, etc.) under a wide range of
conditions (nucleotide concentrations, loading force, neck linker
length and composition, etc.). Besides revealing the mechanism by which
kinesin operates, our model also makes it possible to look into the
experimentally inaccessible details of the mechanochemical cycle and
predict how certain changes in the protein affect its motion.

}

\clearpage

\section*{Introduction}

Conventional kinesin is a microtubule-associated motor
protein which converts chemical energy (stored in ATP molecules) into
mechanical work (by translocating along microtubules towards the
``$+$'' end). The protein is a dimer and uses its two identical motor
domains (heads) alternately to move along microtubules (MTs), in a
manner reminiscent
of walking. Although over the past decades much has been learned about
the structure \citep{Rice:1999bf,Vale:2000mb,hwang_force_2008,Kikkawa:2008uf}
and kinetics \citep{Cross:2004ff} of the individual kinesin heads,
how the motion of two such heads is coordinated during walking is still
poorly understood \citep{Block:2007vo}.
The most plausible hypothesis is that the heads
communicate through a mechanical force mediated by the neck linkers
(NLs, about 13-amino-acid-long peptide chains connecting the heads
and the dimeric coiled-coil tail)
\citep{Rice:1999bf}.
The emerging picture is that the NL can dock to the head 
(\ie, a large section of the NL can bind to and align
with the head domain, pointing towards the forward direction of motion
as demonstrated in Fig.\ {\FIGscheme})
when this head is not in a leading position.
\emph{Neck linker docking}
is, thus, an ideal candidate for both \emph{biasing the diffusion}
\citep{Mather:2006hf} of the
tethered head and \emph{controlling the kinetics} \citep{uemura_loading_2003}
of its head depending
on the position of the other head. The relative importance of these
effects is, however, still debated. The small apparent free energy
change associated with NL docking (both in ATP and ADP
containing heads) \citep{Rice:2003yt} raises, \eg,
the dilemma of whether the docking
of the NL of the bound head is responsible for positioning the
tethered head closer to the forward binding site or, alternatively,
biased forward binding of the tethered head induces passive NL docking in the
other head. Either mechanism can be argued for, if
a particular pathway through the maze of kinetic transitions of the
two-headed kinesin is singled out and modeled with a sufficient number
of parameters.

At present, the only way to test whether NL
docking can indeed provide an adequate explanation for the
operation of kinesin, and to determine how it contributes to force
generation and translocation, is to construct a complete kinetic model
of kinesin that can reproduce all the single molecular experiments
\citep{Visscher:1999qe,Nishiyama:2002hs,Yildiz:2004ms,carter2005mks,Mori:2007zw,Fehr:2008df}
with a single consistent set of parameters.
The challenge is that a realistic kinetic model of kinesin 
requires minimum $6$ different kinetic
states of each head to be distinguished: five of which, the ATP
containing states with the NL undocked ($\sT$) and docked ($\sTs$),
the ADP containing states with the NL undocked ($\sD$) and docked
($\sDs$), and the nucleotide free state with undocked NL ($\sE$),
are MT bound states, while the sixth one is a MT
detached state with ADP in the nucleotide binding pocket ($\sDt$). Many
other states (such as more nucleotide states with docked NL,
alternative MT detached states, several conformational isomers
of the same nucleotide state, a state with both ADP and
P$_\mathrm{i}$ in the nucleotide binding pocket) are also possible,
but as these are either short
lived, never observed experimentally, or both, they can be omitted
without significantly altering the kinetics.
Considering only the $6$ most relevant monomeric states,
kinesin can assume almost $6 \times 6$ different dimeric states
(the actual number is somewhat smaller as one-head-bound
states should only be counted once, while some of the two-head-bound states are
sterically inaccessible), with more than $(2 \times 6) \times (2 \times
6)$ kinetic transitions (as at least $2$ transitions, one
forward and one backward along the chemical cycle, lead out of most
monomeric states). The modeling is further complicated by the
intricate dependence of many of the transitions on both the external
load and the relative positions of the heads.

Here we demonstrate that this seemingly rather complex system
can be treated in a fairly simple and transparent manner. We show that all the
dimeric rate constants (including their load dependence)
can be derived solely from (i) the force-free monomeric
rate constants and equilibrium constants (most
of which are well characterized by direct measurements), and (ii) the
mechanical properties of the NL
(which can be described by the tools of
polymer physics).
The result is a complete, thermodynamically consistent, kinetic model,
which recovers most of the mechanochemical features of
the stepping of kinesin observed to date, but it does so only for a
highly restricted range of parameters. This parameter range is such
that it provides the NL with a crucial role in head coordination.
Both the existence and the uniqueness of a well-functioning parameter
set, as well as the consequent relevance of the docking of the NL
compel us to believe that the model captures the load-dependent
kinetics of kinesin and reveals its walking mechanism
at a level of detail unparalleled in earlier models
\citep{Mather:2006hf,peskin_coordinated_1995,derenyi_kinesin_1996,Liepelt:2007wb}.
A conceptually similar approach was recently taken by Vilfan
\citep{vilfan_elastic_2005} for the description of the motion of another
two-headed motor protein, myosin V.

\section*{Model}

\subsubsection*{Force dependence of the rate constants}
MT bound kinesin heads experience, through their NL, a mechanical force
originating from both the external load and the other head. As this
force depends on the relative position of the heads it can be utilized
to control and coordinate the chemical cycles of the heads. There are,
however, strict thermodynamic constraints on the efficacy of this
control mechanism.
Similarly to the manner in which a kinetic state is constituted by a large
number of microscopic configurations, a kinetic transition between two
states can also be viewed as an ensemble of microscopic trajectories across
the protein's energy landscape.
Thus, an applied force $\vec{f}$ can only affect the rate of a kinetic
transition if it provides an energy contribution along
the microscopic trajectories, in the form of mechanical work.
If, along any such particular microscopic trajectory, the maximal excursion of
the point of application of the force is $\vec{\lambda}$, then the
frequency of that trajectory can only change by at most a factor of
$\exp(\vec{\lambda}\vec{f}/\kT)$, where $\kB$ denotes the Boltzmann
constant and $\temp=293$~K is the absolute temperature.
Consequently, the kinetic transition cannot be sped up or slowed down
by more than this exponential factor.

As the conformation of the kinesin heads is expected to change very
little (considerably less than a nm) during most of the kinetic
transitions or under the typical mechanical forces transmitted by the
NL, and as the magnitude of these forces is of the order of $10$~pN
\citep{Hyeon:2007bk}, the typical work cannot significantly exceed
$\kT\approx 4$~pN~nm. Hence, the corresponding kinetic rate constants
can only be slightly modified by these forces. The only monomeric
transitions that are accompanied by large physical motions (about
$3.5$~nm) are the docking and undocking of the NLs. Therefore, we can
safely neglect the force dependence of any transition except for those
involving NL (un)docking. As these latter processes are expected to be
the key to head coordination and directional bias, we do not attempt to
make any conjecture for the functional form of their force dependence,
but rather we consider the thermodynamics of the NL explicitly using
standard polymer theory.

\subsubsection*{Fast processes}
Kinesin predominantly detaches from the MT when ADP is present in its
nucleotide binding pocket. After one of the heads detaches
(which is necessary for processive stepping), this so-called tethered
head exhibits a diffusive motion within the small confined volume
limited by the length of the NLs. As the diffusion coefficient
of the head is of the order of $10^2$~nm$^2$/$\mu$s, and the
NL length is of the order of $10$~nm, the position of the tethered
head within its diffusion volume equilibrates on the $\mu$s time scale,
much faster than any other rate limiting process during walking. Thus,
the tethered head can always be considered to be in a locally equilibrated
state $\sDt$, in which the probability density or local concentration
of the head is given by the equilibrium mechanical properties of the
NLs (see later).

NL docking, which involves the binding of an about 13
amino-acid-long segment of the NL to the motor core, proceeds
similar to the formation of $\beta$-hairpin structures.
Therefore, it can also
be considered as a fast process, and the docked and undocked
configurations can be treated as being in local equilibrium at any moment.
Taking this into account the kinetic model can be further simplified by
introducing the compound states $\sTTs$ (by merging $\sTs$ with $\sT$)
and $\sDDs$ (by merging $\sDs$ with $\sD$), representing MT
bound kinesin heads having, respectively, ATP or ADP in their
nucleotide binding pocket, irrespective of the configuration of the
NL. Since the kinetic transitions from the two elementary states of a
compound state can be different (see, \eg, Fig.\ {\FIGscheme}a), it is
important to note that their relative frequencies can always be
recovered from their free energy difference which, obviously, depends
on the state and position of the other head, and also on the applied
external load.

\subsubsection*{Two-dimensional state space}
The most natural way of visualizing the kinetics of a kinesin dimer is
to arrange the dimeric states into a two-dimensional lattice, as shown
in Fig.\ {\FIGtwod}, where the horizontal direction represents both
the location and the state of one of the heads, whereas the vertical
direction represents the same for the other head. Since kinesin walks
primarily along a single protofilament (rarely stepping sideways)
\citep{ray_kinesin_1993,yildiz2008isc}
with a hand-over-hand mechanism
\citep{Yildiz:2004ms,asbury_kinesin_2003,schief_inhibition_2004},
one of the heads (represented
horizontally) can only be bound to the odd-numbered $\beta$-tubulins
(if $\beta$-tubulins are numbered along the protofilament, ascending
towards the ``$+$'' end of the MT), while the other head can
bind to either of the two neighboring even-numbered
$\beta$-tubulins. Therefore, only lattice points near the diagonal of
the state space (where the distance between the two heads is not larger
than the $L \approx 8$~nm periodicity of the protofilament), marked by
solid black squares, are allowed.

For practical convenience each $\sDDs$ state is split into $\sDDst$ and
$\sDDsl$, where the subscripts ``t'' and ``l'', respectively, indicate
that the head is in a trailing or a leading position (\ie, closer to
the ``$-$'' or ``$+$'' end of the MT). This split has the benefit
of flattening the kinetic pathways, as the sequence of monomeric
states (\dots, $\sDt$, $\sDDsl$, $\sE$, $\sTTs$, $\sDDst$, $\sDt$,
\dots) along each axis (from left to right and from bottom to top) reflects
the succession of the states of each head during ``standard'' forward
walking: the tethered head ($\sDt$) binds forward to the next
$\beta$-tubulin, becoming a leading head ($\sDDsl$); releases
its ADP ($\sE$); binds a new ATP ($\sTTs$); and by the time the ATP is
hydrolyzed into ADP, the other head will have stepped
forward, leaving this head in a trailing position ($\sDDst$); from which
it eventually detaches from the MT ($\sDt$) to begin a new
cycle. As the subscript of $\sDDs$ uniquely specifies the relative
positions of the two heads in the two-head-bound states, the lattice points
that do not conform with this geometry are removed from the set of
allowed states (gray shading in Fig.\ {\FIGtwod}). The only ambiguity
occurs for one-head-bound states, when the bound head is in the $\sDDs$
state, because it cannot be designated as either trailing or leading.
As a remedy, we artificially assign $\sDDsl$ to such
$\sDDs$ states, and disregard the corresponding $\sDDst$ lattice points
(indicated by black crosses). This way
the allowed lattice points can be grouped into $4 \times 3$
rectangular blocks, each being composed of a $3 \times 3$ array of
two-head-bound states and a $1 \times 3$ array of one-head-bound
states. Due to the equivalence (permutation invariance) of the two
heads all these blocks are identical, with every other block being
mirrored about the diagonal. Advancing from one block to a neighboring
one corresponds to kinesin taking a step. There is also one
special lattice point near each block, the ($\sDt$,$\sDt$) point
denoted by an open square, representing a kinesin molecule with both of
its heads detached from the MT, which can be viewed as the
source and sink (or initial and final stages) of walking.

After setting up the state space, the next step is to identify all the
possible kinetic transitions between the dimeric states and to
determine their rate constants (in both directions). By construction,
all horizontal and vertical transitions between neighboring lattice
points (termed lattice transitions, indicated along the axes of
Fig.\ {\FIGtwod} by straight double arrows: $\tr$)
certainly exist. There are also some oblique non-lattice transitions
between points ($\sDt$,$\sDDsl$) and ($\sDDsl$,$\sDDst$), and also
between their mirror images, marked by straight double arrows, which
are inherited from the lattice transitions involving the disregarded
one-head-bound states (marked by crosses).
And finally, some horizontal and vertical non-lattice
transitions, indicated by curved double arrows along the axes, can also
exist, which correspond to futile ATP hydrolysis, $\sTTs \tr
\sDDsl$ (and its reverse), as well as to the release of ADP by the
trailing head (also resulting in a futile ATP hydrolysis), $\sDDst \tr
\sE$ (and its reverse).

To simplify notation and to treat each dimeric state with its
counterpart in the mirrored block together, in the following we will
denote each dimeric state as $AB$, where $A$ and $B$, respectively, stand
for the monomeric states of the trailing and leading heads for a
two-head-bound construct, and of the MT bound and tethered heads for a
one-head-bound construct. The kinetic rate constant of a transition
from either a monomeric or dimeric state $a$ to state $b$ will be
denoted as $k_{a \to b}$, and the corresponding equilibrium constant as
\beq
K_{a,b} = \frac{k_{a \to b}}{k_{b \to a}} =
\e^{-\frac{\Delta G_{a,b}}{\kT}} ,
\eeq
where $\Delta G_{a,b}=G_b-G_a$ is the free energy difference between
the two states.

\subsubsection*{Thermodynamic consistency}
Thermodynamics requires that along any closed series of subsequent
transitions (which is often referred to as a thermodynamic box) the
product of the equilibrium constants be
\beq
K_{a,b} K_{b,c} \cdots K_{z,a} =
\e^{-n\frac{\Delta G_\mathrm{ATP}}{\kT}} ,
\label{eq:th_box}
\eeq
where $n$ denotes the number ATP molecules hydrolyzed along one sequence
of transitions (with $n<0$ corresponding to ATP synthesis) and
\beq
\Delta G_\mathrm{ATP} = \Delta G^0_\mathrm{ATP}
- \kT \ln\left(\frac{[\mathrm{ATP}]}{[\mathrm{P_i}][\mathrm{ADP}]}\right)
\label{eq:ATP}
\eeq
is the free energy change of ATP hydrolysis (with
$\Delta G^0_\mathrm{ATP} \approx -30.5$~kJ/mol $\approx -12.5 \kT$ being
the standard free energy change and the square brackets denoting
concentration). Unless otherwise noted the values $[\mathrm{P_i}]=1$~mM and
$[\mathrm{ADP}]=0.01[\mathrm{ATP}]$ are assumed.

Due to the equivalence of the blocks of the two-dimensional state space
(originating from the periodicity of the MT) the above relation and also the
notion of the thermodynamic box can be generalized to any series of
subsequent transitions that starts at an arbitrary dimeric state $a$
and ends at an identical state $a'$ that is $m$ periods forward along
the MT:
\beq
K_{a,b} K_{b,c} \cdots K_{z,a'} =
\e^{- \left( n\frac{\Delta G_\mathrm{ATP}}{\kT} - m F L \right) } ,
\label{eq:gen_th_box}
\eeq
where $F$ is the longitudinal (\ie, parallel to the direction of
forward walking) component of the external force exerted on the kinesin
and, thus, $- m F L$ is the work done by the kinesin on the external
force during $m$ forward steps.

The thermodynamic boxes and their generalized versions can be used for
either verifying that the calculated rate constants are indeed
consistent with the laws of thermodynamics or determining certain
equilibrium constants and rate constants that are otherwise unknown or
difficult to deduce from microscopic considerations.

\subsubsection*{Dimeric rate constants}
All dimeric rate constants and equilibrium
constants under arbitrary external load can be derived from the monomeric
rate constants (listed in Table~{\TABparameters}), the free energy
changes of NL docking under zero force ($\Delta G_{\sT,\sTs} = - \kT
\ln K_{\sT,\sTs}$ and $\Delta G_{\sD,\sDs} = - \kT \ln K_{\sD,\sDs}$,
also shown in Table~{\TABparameters}), and the mechanical
properties of the NL. Due to thermodynamic consistency some of the
monomeric rate constants, such as the ATP synthesis rate constants
in both the NL undocked and docked configurations ($k_{\sD \to \sT}$
and $k_{\sDs \to \sTs}$), cannot be set independently and should be
determined from the corresponding thermodynamic boxes ($\sE \tr \sT \tr
\sD \tr \sE$ and $\sE \tr \sT \tr \sTs \tr \sDs \tr \sD \tr \sE$).
Similarly, the reverse rate constants of nucleotide release from the NL docked
configurations ($k_{\sE \to \sTs}$ and $k_{\sE \to \sDs}$) have to be
determined from thermodynamic boxes ($\sE \tr \sT \tr \sTs \tr \sE$ and
$\sE \tr \sD \tr \sDs \tr \sE$).

The dimeric transitions can be classified into two groups based on
their impact on the NL. One of the groups consists of all the
transitions that are accompanied by the configurational change of the
NL, either through MT binding/unbinding of the head (such as
$\sT\sDt \tr \sT\sD$, $\sT\sDt \tr \sDs\sT$, \etc.) or the
docking/undocking of the NL (such as $\sT\sD \tr \sTs\sD$, $\sT\sDt \tr
\sTs\sDt$, \etc.). These are the transitions that depend both on the
magnitude and direction of the external force as well as on the states and
relative positions of the two heads and, therefore, require the careful
consideration of the dynamics of the NL. The rest of the transitions,
which constitute the second group (such as the uptake/release of
ATP/ADP with undocked NL, or the hydrolysis/synthesis of ATP), have no
such force and position dependence, and their rate constants are considered
identical to those of their force-free monomeric counterparts.

\subsubsection*{NL dynamics}
As the undocked NL (and also the unbound fragment of the docked NL) is
thought to assume a random coil configuration with a persistence length
($\Lp$) in the range of $0.4-0.5$~nm \citep{Hyeon:2007bk},
practically any polymer model
(as long as it respects the persistence length and it does not let the
polymer stretch beyond its contour length) can be used to describe its
equilibrium mechanical properties. We have chosen the
freely jointed chain (FJC) model, because it conveniently allows the
independent treatment of the connected segments of the two NLs. We 
further simplified the mechanical model by neglecting any non-specific
interaction and steric repulsion between the heads, the NLs, and the
MT, because we believe that these are subordinate to the effects of the
docking enthalpy and the configurational entropy of the NL, and also
because we intend to keep the model as simple and free of unimportant
details as possible to demonstrate its predictive power.

The Cartesian coordinate system is chosen such
that its $x$ axis runs parallel to the protofilaments of the MT
pointing towards the ``$+$'' end, the $y$ axis points perpendicularly
away from the MT surface, and the $z$ axis is perpendicular to both and
tangential to the MT surface. Each NL is built up of $N=\Nd+\Nu$ Kuhn
segments (or bonds) of length $\LK=2\Lp$, out of which only the first
$\Nd$ take part in the docking by aligning along the head in the $x$
direction as represented by the vector $\vLd=(\Ld,0,0)$,
while the last $\Nu$ remain
undocked. For these geometric parameters the following values are
assumed: $\Lp=0.46$~nm, $\Nd=4$, $\Nu=1$, and $\Ld=3.5$~nm, which
are compatible with the real structural properties of the heads and the
NLs. Note that with these values the leading head is unable to dock its
NL, which slightly reduces the number of attainable states of kinesin
and somewhat simplifies the overall kinetic scheme.

The external force $\vF=(F,|F|\tan\alpha,0)$, where the negative of the
lateral component ($-F$) is conventionally referred to as the load,
is applied to the joint of the two NLs via the coiled-coil
tail of the kinesin. The angle $\alpha$ of the force to the MT depends
on the details of the experimental setup, in particular, on the length
of the coiled-coil tail and the size of the bead in the optical trap.
Throughout the paper we use a reasonable value of $\alpha=45^\circ$,
although the results do not change much as long as $\alpha$ stays below
about $60^\circ$.

The FJC model readily provides the probability density
$\rho_{N}^{0}(\vR)$ of the end-to-end vector $\vR$ of a random polymer
chain of $N$ Kuhn segments (for details see Ref.\ \citep{czovek2008rnl}).
In any one-head-bound state the convolution of the probability
densities of the undocked segments of the two NLs combined with the
appropriate Boltzmann weights gives then the local concentration of the
starting (N-terminal) point of the NL of the tethered head measured from the
starting point of the NL of the bound head:
\beq
c(\vR,\vF) = \frac{ \int
 \rho_{N}^{0}(\vR') \e^{\frac{\vF\vR'}{\kT}}
 \rho_{N}^{0}(\vR-\vR')
 \dd\vR' }{ \int
 \rho_{N}^{0}(\vR') \e^{\frac{\vF\vR'}{\kT}}
 \dd\vR' }
\label{eq:c}
\eeq
if the bound head's NL is undocked and
\beq
c^{*}(\vR,\vF) = \frac{ \int
 \rho_{\Nu}^{0}(\vR'-\vLd) \e^{\frac{\vF\vR'}{\kT}}
 \rho_{N}^{0}(\vR-\vR')
 \dd\vR' }{ \int
 \rho_{\Nu}^{0}(\vR'-\vLd) \e^{\frac{\vF\vR'}{\kT}}
 \dd\vR' }
\label{eq:cs}
\eeq
if it is docked. These local concentrations can also be viewed as good
approximations of the local concentrations of the tethered head during
its diffusive motion. Multiplying them by the second order binding rate
constant $k_{\sDt\to\sD}$ at the forward and backward binding positions
($\vL=(L,0,0)$ and $(-\vL)$, respectively) will thus yield the force
dependent dimeric rate constants from any tethered state to the
corresponding two-head-bound state. Unbinding is always considered to
occur with the monomeric rate constant $k_{\sD\to\sDt}$.

The FJC probability density can also be used to express the equilibrium
constants between the undocked and docked NL configurations of the
monomeric (or dimeric one-head-bound) compound states under external
force:
\beq
K_{A,A^{*}}(\vF) =
 \e^{-\frac{\Delta G_{A,A^{*}}}{\kT}}
 \frac{ \int
 \rho_{\Nu}^{0}(\vR'-\vLd) \e^{\frac{\vF\vR'}{\kT}}
 \dd\vR' }{ \int
 \rho_{N}^{0}(\vR') \e^{\frac{\vF\vR'}{\kT}}
 \dd\vR' }
\label{eq:KsTTs}
\eeq
with $K_{A \sDt, A^{*} \sDt}(\vF) = K_{A,A^{*}}(\vF)$ and
$A$ standing for either $\sT$ or $\sD$.

There are three more types of dimeric transitions (exemplified by the
three dashed double arrows in Fig.\ {\FIGscheme}) that are
accompanied by NL configurational change, but these are very difficult
to characterize directly by means of microscopic polymer dynamics. Each
of them, however, is a part of a thermodynamic box, in which all the
other transitions are known or computable and, therefore, can be
characterized by closing the thermodynamic box.

The first type is the docking/undocking of the NL by the trailing head
within a two-head-bound compound state (demonstrated by the cartoon and
kinetic scheme in Fig.\ {\FIGscheme}a). Their equilibrium constants can
be summarized as
\beq
K_{A B, A^{*} B}(\vF) =
K_{A,A^{*}}(\vF)
 \frac{c^{*}(\vL,\vF)}{c(\vL,\vF)} ,
\label{eq:AAs}
\eeq
where
$A$ stands for $\sT$ or $\sD$, and
$B$ for either $\sT$, $\sD$, or $\sE$.

The second type is the binding of the tethered head to a backward
binding site with the NL in the docked configuration (Fig.\ {\FIGscheme}b):
\beq
k_{B \sDt \to \sDs B}(\vF) =
k_{\sDs \to \sDt}
 \frac{k_{\sDt \to \sD} c(-\vL,\vF)}{k_{\sD \to \sDt}}
 K_{\sD B,\sDs B}(\vF) ,
\label{eq:DtDs}
\eeq
where again $B$ can be either $\sT$, $\sD$, or $\sE$.

The third type is the uptake of a nucleotide by an empty MT-bound head
with a simultaneous docking of its NL (as in Fig.\ {\FIGscheme}c):
\beq
k_{\sE C \to A^{*} C}(\vF) =
k_{A^{*} \to \sE}
 \frac{k_{\sE \to A }}{k_{A \to \sE}}
 K_{A C, A^{*} C}(\vF) ,
\label{eq:EAs}
\eeq
where
$A$ stands for $\sT$ or $\sD$, and
$C$ for either $\sT$, $\sD$, $\sE$, or $\sDt$.

\subsubsection*{Parameter fitting}
Using the full set of kinetic rate constants between the dimeric states
and equilibrium constants within the compound states obtained in the
above manner, one can (i) solve the kinetic equations for the
steady-state occupancies and kinetic fluxes exactly to determine some
of the simplest average characteristics of kinesin's movement (such as
its velocity, ATP-hydrolysis rate, processivity, \etc.); and also (ii)
perform kinetic Monte-Carlo simulations to generate \textit{in-silico}
trajectories and to deduce more complicated quantities (such as
frequencies and dwell times of forward and backwards steps separately,
randomness, \etc) under various experimental conditions for arbitrary
model parameters. Most parameter sets, however, result in unrealistic
behavior for kinesin. In order to find parameters for which the model
reproduces the experimentally observed behavior, we prescribed $10$
different criteria taken from the literature \citep{carter2005mks}
(including the average velocity, processivity, hydrolysis rate, ratio
of forward and backward steps at specific ATP concentrations and loads,
as well as the stall load; see Table~S{\TABcriteria}
in the Supporting Material for details) and
performed a \emph{simulated annealing} optimization in the space of the
kinetic parameters
\colorA{(as listed in the ``optimal range'' column of Table~{\TABparameters}).}
As the
geometric parameters of the NL are highly constrained, we omitted them
from the optimization. We found that regardless of where the
optimization starts from, the parameters always end up in a very narrow
range (Table~{\TABparameters}), within which all criteria are  
satisfied simultaneously with good accuracy. To achieve this,
however, we also had to introduce a slow $\sT\to\sDt$ transition (with
its reverse determined from a thermodynamic box), otherwise the
backward steps at very high loads ($>10$~pN) would have taken too long.
\colorA{The small value of this parameter, however, ensures that it has
negligible effects under normal loading conditions.}
In Table~S{\TABdeviations} we demonstrate how the deviation of the
model parameters from their optimal values affects some of the most
relevant experimental observables of kinesin.

\section*{Discussion}

Remarkably, the narrow parameter range obtained by the optimization is
highly consistent with the experimental values (with some deviation for
the NL docking free energies, discussed later), as shown in
Table~{\TABparameters}.
\colorA{One could speculate that if the performance of kinesin had long
been under evolutionary pressure, then not much room must have been
left for the values of the kinetic parameters that could result in the
same observed behavior.}
%
The fact that our optimization has resulted in a practically identical
and similarly constrained parameter set is a strong justification of
the credibility of our model.
Moreover, the optimal parameter range not only allows the model to
satisfy the prescribed $10$ criteria, but also to reproduce the vast
majority of the available experimental data reasonably well.
To demonstrate this, we have replicated some of the best known and
highest quality experiments using our kinesin model, with a fixed set
of model parameters (see Table~{\TABparameters}) selected from the
optimal range. Our model can reproduce the load vs.\ dwell time
curves by Carter and Cross \citep{carter2005mks}, for both forward and
backward steps at saturating
($1$~mM) and low ($10$~$\mu$M) ATP concentrations under the full range
of external load between $-15$ and $15$~pN (see Fig.\ {\FIGcarter}).

The ratio of the numbers of forward and backward steps for both large
loading and assisting forces converges to exponential functions with
the force constant of $\kT/L$ (indicated as dotted lines in
Fig.\ {\FIGcarter}), as expected from the exponential decline of the
local concentration of the tethered head near the unfavorable binding
site. The transition between the two limiting exponentials seems to
follow a less steep exponential with a force constant of approximately
half the magnitude, also in reasonable agreement with the experimental data
at both ATP concentrations.
\colorB{Note, however, that full agreement is limited by the differences
in the methods of step detection. Experimentally the trajectories of a
bead in an optical trap are analyzed (where, \eg, short backward steps
can easily be mistaken for bead fluctuations or vice versa), whereas in
our model we define a step as the arrival of kinesin at a
one-head-bound state from a neighboring one-head-bound state (as the
complete dynamics of an attached bead cannot be considered at this
level of modeling). The two methods might, thus, result in slightly
different step counts (with little or no effect on any other
observables).}

Fig.\ {\FIGcarter} also demonstrates that as long as the external
load is smaller than the stall load (about $6-7$~pN) by at least a few
pN's, the number of ATP molecules hydrolyzed per step is close to one
\citep{Visscher:1999qe,schnitzer1997kho} and the processivity of kinesin
is over $100$ steps \citep{Vale:1996pc,yajima2002dlo}.
\colorB{Such a high processivity at low load can be achieved, because
kinesin (with the parameters in Table~{\TABparameters}) has only about
a 10\% chance of getting into the $\sDDs\sDt$ state from the
$\sTTs\sDDs$ state (by ATP hydrolysis and MT detachment by the trailing
head), and another 10\% chance from the $\sTTs\sDt$ state (by ATP
hydrolysis). From the one-head-bound $\sDDs\sDt$ state, however, the
bound head can rapidly release its ADP, and it has only about a 5\%
chance of detaching from the MT and ending the processive motion
instead. Thus, the total chance of two head detachment per step (which
is the product of the 20\% and the 5\%) is about 1\%. For increased
loading force the rate of forward binding from the $\sTTs\sDt$ state
decreases, which increases the chance of getting into the $\sDDs\sDt$
state and decreases the processivity (simultaneously with the
velocity). For assisting forces, on the other hand, the rate of forward
binding from the $\sTTs\sDt$ state increases, thereby increasing the
processivity.}

Another important set of experimental data concerns the average
velocity and the randomness of the stepping of kinesin measured by
Block et al.\ \citep{schnitzer1997kho,visscher1999skm}. Our model
reproduces these data with good accuracy both as functions of the
ATP concentration and the external load (see Fig.\ {\FIGATPvelrand}).
\colorA{At no load and high ATP concentrations the three kinetic rate
constants that limit the velocity of kinesin and lead to low randomness
can clearly be identified as $k_{\sTs\to\sDs}$, $k_{\sDs\to\sDt}$, and
$k_{\sD \to\sE }$ in Table~{\TABparameters}.}

A more profound test of the validity of the model is, however, when one tries
to reproduce the behavior of kinesin under highly non-physiological
conditions. Yildiz et al.\ \citep{yildiz2008isc} recently set the ATP
concentration to zero, and then applied a $1$~pN assisting and a $2$~pN
loading force to kinesin at several ADP concentrations. Even in this
extreme situation, when the steps were initiated by ADP uptake, the
results of our simulations (Fig.\ {\FIGmichio}) show very good
agreement with the experimental data. The same authors also elongated
the NL of kinesin by the insertion of $14$ amino-acid-long
glycine-serine repeats, and observed that the velocity of the motor
dropped down significantly at zero force, but as the assisting force
was raised above $6$~pN the velocity exceeded even that of the wild
type. Our results
\colorB{(by raising the number of undocking Kuhn segments of the NLs
from $\Nu=1$ to $\Nu=6$)}
indeed show a similar drop at zero force, and an
increasing velocity for increasing assisting force, although at a smaller pace
(see also Fig.\ {\FIGmichio}).
The reason for this discrepancy might be that either the $y$ component
of the pulling force in the experiments is smaller or there is some sort
of nonspecific attraction between the MT and some part (head/NL/tail) of
kinesin.

The NL of kinesin has also been modified by either a partial or a complete
replacement of its amino acid sequence \citep{Rice:2003yt,case2000rkn}. In our
approach this can be taken into account by increasing the free energy
changes of NL docking (simultaneously for $\Delta G_{\sT,\sTs}$ and
$\Delta G_{\sD,\sDs}$) 
The predictions of our model (Fig.\ S{\FIGG}) are again in very good
agreement with the experiments: increasing $\Delta G_{\sT,\sTs}$ and
$\Delta G_{\sD,\sDs}$ up to $12$~$\kT$ results in a slowly decreasing
stall load with a rapidly decreasing velocity at zero load
\citep{Rice:2003yt,case2000rkn}; a slowly decreasing processivity
\citep{Rice:2003yt}; and a slightly changing ATPase activity
\citep{case2000rkn}.
For an even more drastic, $24$~$\kT$ increase of the docking free
energies (which is practically equivalent to prohibiting the docking of
the NL), the walking capability of kinesin diminishes, supporting
the importance of NL docking in the motility of kinesin.

\colorB{Our model is also consistent with the half-site reactivity
experiments by Hackney \citep{hackney1994eah}, because upon the first
contact of a kinesin (containing an ADP in each head) with the MT, only
one of the heads is able to bind to the MT and release its ADP rapidly.
As this MT bound empty head keeps its NL undocked, the other head has a
very low local concentration at the nearest binding sites, therefore,
its MT binding and ADP release rates become very low.}

The only parameters for which the optimal range deviates noticeably
from the experimental values are the free energy changes of NL docking:
the $\Delta G_{\sT,\sTs}$ range stays below, whereas the $\Delta
G_{\sD,\sDs}$ range lies above the values measured by Rice et al.\
\citep{Rice:2003yt}. However, the consistency of our model with the
broad variety of single molecular mechanical studies provide a strong
support in favor of our predicted values, and demands for an
experimental reexamination of these parameters. Our studies indicate
that the optimal range can be shifted closer to the measured values
only if the 6.75~pN constraint on the stall load is lowered (see
Fig.\ S{\FIGgtgd}). Molecular dynamics simulations are also
consistent with a larger free energy difference between NL docking in
the ADP and ATP state of the head \citep{hwang_force_2008}. Possible
sources of error in the original experiments \citep{Rice:2003yt} might
be the use of an ATP-analogue, the influence of the spin labels, or the
spin labels not reporting on the strong stabilizing binding of the few
last amino acids of the NL to the motor domain \citep{hwang_force_2008}.
Nevertheless, even our predicted value for $\Delta G_{\sT,\sTs}$ is far
from being sufficient to explain a pure \emph{power stroke} mechanism.
\colorA{Therefore, other mechanisms, such as position dependent MT
binding/unbinding of the head (called \emph{biased capturing}),
are clearly at play and employed by kinesin.}

The values of some of the rate constants in Table~{\TABparameters}
reveal how the NLs play the role of position sensors and carry out
the coordination of the kinetics of the heads. First, if an ADP
containing head is in a trailing position, then its NL is forced into
its docked configuration. Thus, the relation
$k_{\sDs\to\sE} \ll k_{\sDs\to\sDt}$ ensures that the trailing head
rapidly detaches from the MT before releasing its ADP. Conversely,
after the diffusing head binds to the MT in a leading position,
where NL docking is sterically inhibited, the relation
$k_{\sD\to\sE} \gg k_{\sD\to\sDt}$ ensures the fast release of ADP,
resulting in strong MT binding. Similarly, whenever an ATP containing
head is in a leading position, the relation
$k_{\sT\to\sE} \gg k_{\sT\to\sD}$ prevents the head from prematurely
hydrolyzing its ATP by favoring its release. However, as soon as this head
becomes trailing (due to forward binding of the other head), the relation
$k_{\sTs\to\sE} \ll k_{\sTs\to\sDs}$ accelerates ATP hydrolysis.
The strong dependence of some of the rate constants on the state
of the NL (often referred to as D- and T-gates \citep{Block:2007vo})
allows the
kinesin to efficiently avoid futile ATP hydrolysis and to keep the
kinetics of its heads in synchrony.

\colorB{Our model thus not only recovers the existence of the main gating
mechanisms, but it also provides a detailed explanation for their
physical origin: The D-gate of the trailing head (\ie, its preference
for MT detachment rather than ADP release) is the consequence of the
tension in the NLs, which forces the NL of the trailing head into the
docked configuration, thereby accelerating its MT detachment and
slowing down its ADP release. The T-gate of the leading head (\ie, its
strongly reduced ATPase activity) is also ensured by the tension in the
NLs, which forces the NL of the leading head into the undocked
configuration, where the binding of an ATP is quickly followed by the
release of the same ATP molecule, thereby preventing its hydrolysis in
most of the time.}

The NL configuration dependent rate constants also explain the
observed dependence of the ADP and MT affinity of the heads on the
direction of pulling \citep{uemura_loading_2003}. Although a much weaker
strain dependence of some other transitions (not considered in our
model) cannot be ruled out, the main factor in head coordination seems
to be the docking/undocking of the NL.

In conclusion, by considering only the force-free rate
constants and free energy changes of monomeric kinesin,
combined with the basic mechanical
properties of the NL, we were able to construct a kinetic model that
reproduces practically all the mechanochemical features of the
stepping of kinesin.
\colorA{This was achieved by (i) collecting all the possibly relevant
kinetic states and mechanical properties of the monomers, (ii) putting
them together into a complete kinetic model using thermodynamics as the
only constraint, (iii) and letting the model find its parameters by
prescribing a diverse set of criteria deduced experimentally.}
\colorA{The fact that a narrow parameter range (in
agreement with the values from the literature) has been found implies
that the initial assumptions about the relevant states and properties
of the heads are sufficient, and the model can reproduce the behavior
of kinesin in a detailed and realistic manner, with immense relevance
in planning and interpreting experiments.}
The obtained model is thus complete both kinetically (as
all the possible transitions are considered) and thermodynamically (as
all the thermodynamic boxes are closed).
To demonstrate that the complete kinetics in the two-dimensional
state space is indeed necessary for the modeling of kinesin we have
prepared two movies (Movies S1 and S2),
which show that the steady-state fluxes are not
concentrated along any specific pathway and that the flux
distribution is very sensitive to both the ATP concentration and the
external load. Our model can also be viewed as a general framework for
testing various hypotheses, as it can be implemented easily, its
parameters can be modified at will, and it can be conveniently extended
to embrace more kinetic states (including other NL configurations)
and intermolecular interactions. We also provide a web site
[http://kinesin.elte.hu/] where it is possible to run simulations and to test
the model with arbitrary parameters.

\subsubsection*{Acknowledgments}

This work was supported by
the Hungarian Science Foundation (K60665)
and the Human Frontier Science Program (RGY62/2006).
The authors are grateful to
M. Kikkawa and M. Tomishige
for helpful discussions.

\bibliography{articles}

\clearpage

\begin{table}
\caption{Monomeric rate constants and free energy changes}
\delimwidth
\begin{tabular}{p{45pt}p{30pt}p{51pt}p{45pt}p{51pt}p{75pt}}
\\
parameter & model value & optimal range & values in literature & unit & Refs.\vphantom{
} \\ \hline
$\Delta G_{\sT,\sTs}$ & -7  & -8-- -4   & $\sim$ -1         & $\kT$           & \citep{Rice:2003yt} \\
$\Delta G_{\sD,\sDs}$ & 5.5 & 3--10     & $\sim$ 1          & $\kT$           & \citep{Rice:2003yt} \\
$k_{\sT \to\sE }$     & 100 & 40--120   & \mrd{1--300$^*$}  & \mr{s\rec}      & \mr{\citep{ma1995kmk,gilbert1994pss,gilbert1995ppa,Farrell:2002rv}} \\
$k_{\sTs\to\sE }$     & 0   & 0--0.01   &                   &                 &  \\
$k_{\sE \to\sT }$     & 3.8 & 2--4      & 1--6              & s\rec$\mu$M\rec & \citep{ma1995kmk,gilbert1994pss,gilbert1995ppa,Farrell:2002rv,auerbach2005asa} \\
$k_{\sD \to\sE }$     & 300 & 90--1000  & \mrd{10--1000$^*$}& \mr{s\rec}      & \mr{\citep{ma1995kmk,gilbert1994pss,gilbert1995ppa,Farrell:2002rv,auerbach2005asa,ma1997kmm,ma1997ihm,rice1999sck}} \\
$k_{\sDs\to\sE }$     & 0   & 0--50     &                   &                 &   \\
$k_{\sE \to\sD }$     & 1.5 & 0--5      & 1.5               & s\rec$\mu$M\rec & \citep{ma1997ihm} \\
$k_{\sT \to\sD }$     & 10  & 0--40     & \mrd{70--500$^*$} & \mr{s\rec}      & \mr{\citep{gilbert1994pss,gilbert1995ppa,Farrell:2002rv,auerbach2005asa}} \\
$k_{\sTs\to\sDs}$     & 200 & 60--200   &                   &                 &   \\
$k_{\sD \to\sDt}$     & 8   & 0--100    & \mrd{10--100$^*$} & \mr{s\rec}      & \mr{\citep{ma1995kmk,auerbach2005asa,ma1997kmm,crevel2004kdr}} \\
$k_{\sDs\to\sDt}$     & 105 & 100--1000 &                   &                 &   \\
$k_{\sDt\to\sD }$     & 20  & 5--40     & 10--20            & s\rec$\mu$M\rec & \citep{gilbert1995ppa,moyer1998pah} \\
$k_{\sT \to\sDt}$     & 3   & 1--4      & N/A               & s\rec           & N/A \\ \hline
\end{tabular}
\\
\centering $^*$ NL configuration was not resolved experimentally
\label{tab:parameters}
\end{table}

\clearpage
\section*{Figure Legends}

\begin{figure}
\centerline{\includegraphics[width=3.5in]{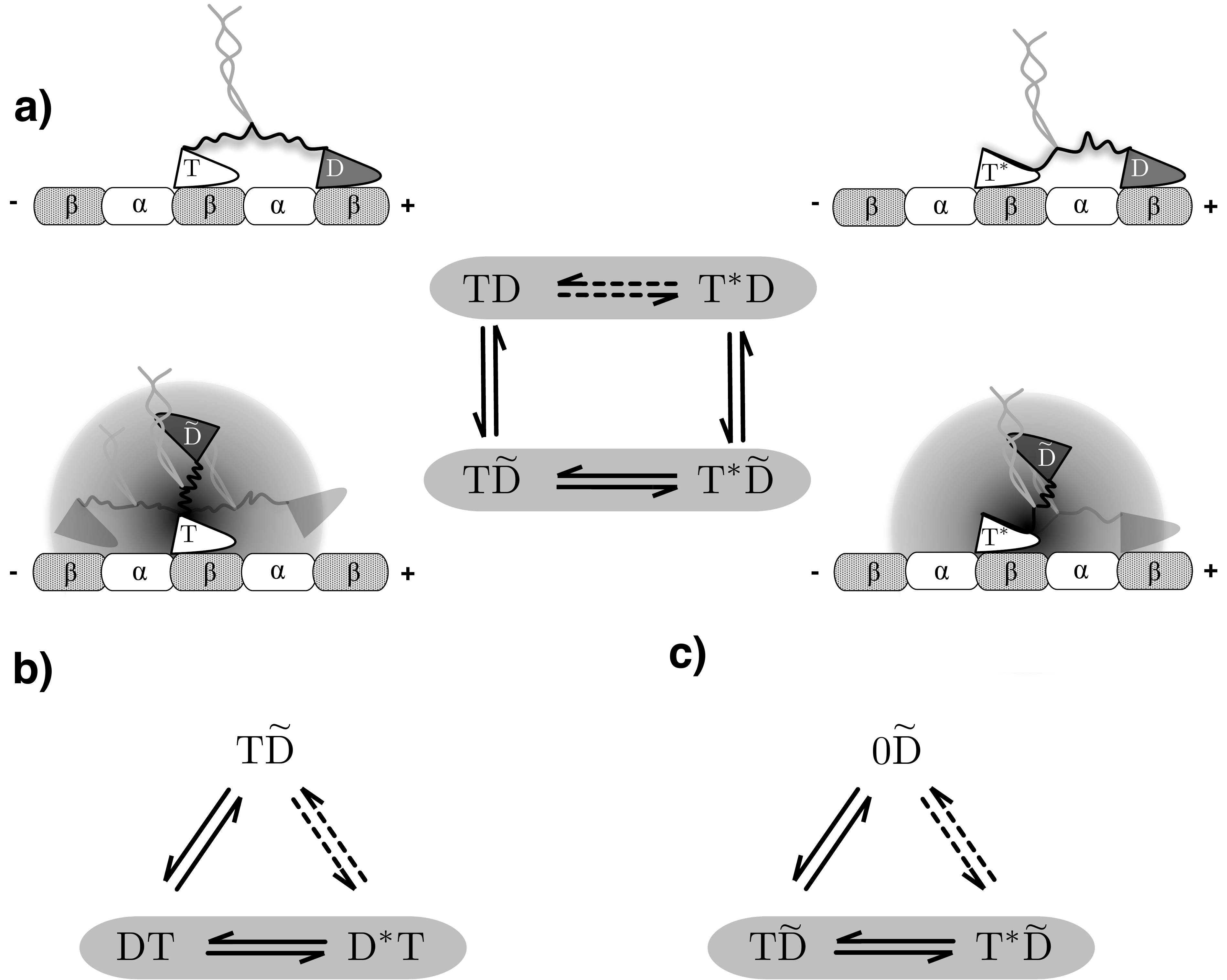}}
\caption{}
\label{fig:scheme}
\end{figure}

\subsubsection*{Figure~\ref{fig:scheme}.}
Neck linker docking scheme.
(a) The cartoons illustrate the geometries of the one-head-bound and
two-head-bound states of kinesin with both docked and undocked neck
linkers. The thermodynamic box corresponding to the cartoons is
depicted in the middle. (a), (b), and (c) show examples for the three
basic types of thermodynamic boxes that occur in the model.
Each box is used to determine the equilibrium constant of one
of the transitions (dashed double arrows).

\begin{figure}
\centerline{\includegraphics[width=3.5in]{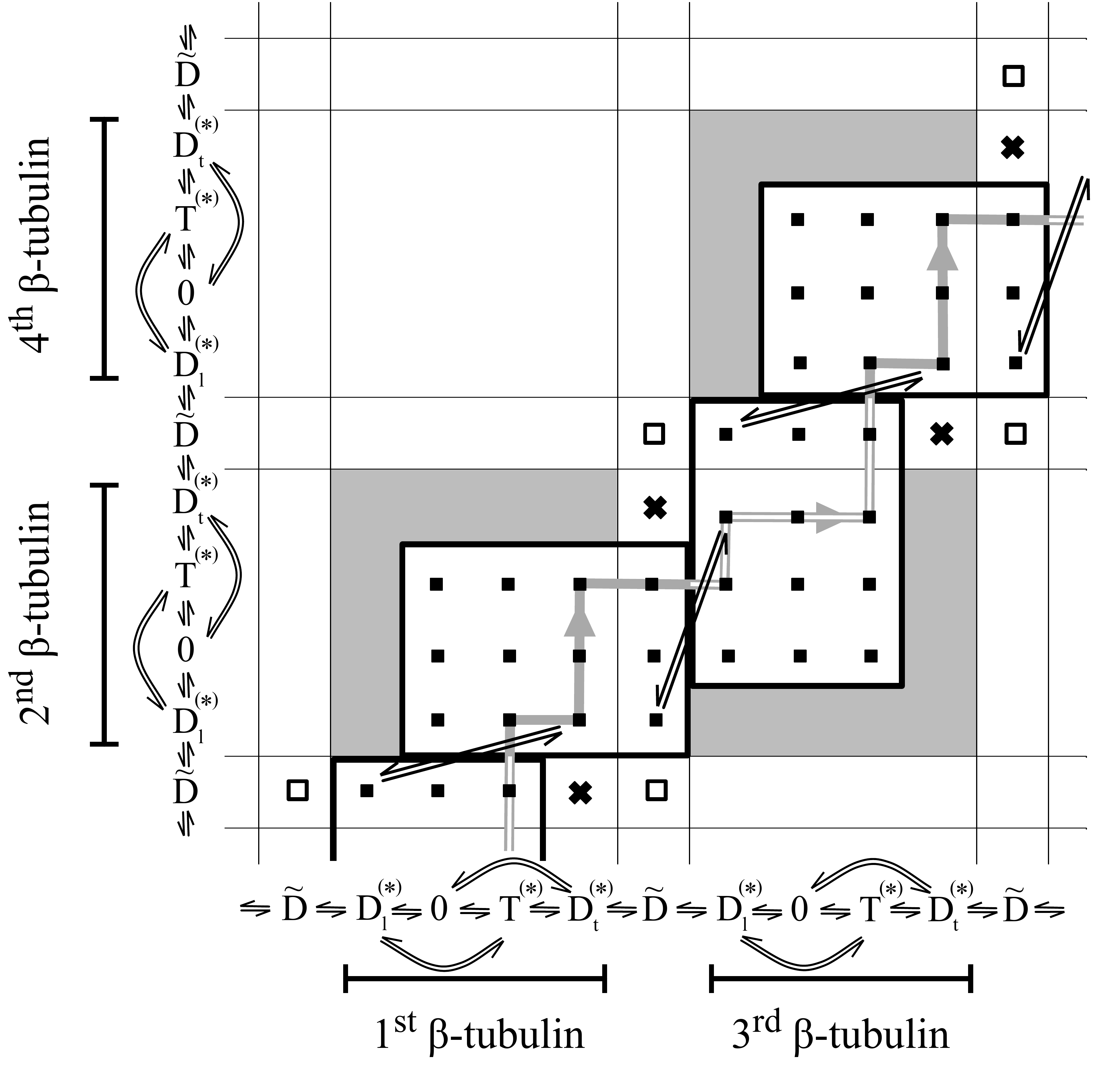}}
\caption{}
\label{fig:2d}
\end{figure}

\subsubsection*{Figure~\ref{fig:2d}.}
Two-dimensional state space of dimeric kinesin.
Each axis represents both the location and the state of one of the
heads. The subscripts ``t'' and ``l'' explicitly refer to the trailing
and leading positions of the head. Allowed MT-bound states are marked by
solid black squares, and the detached states are marked by open squares.
The crosses indicate that the trailing positions in the one-head-bound
states are disregarded (in favor of the leading positions). The possible
kinetic transitions are denoted by double arrows, either along the axes
or inside the state space. The most typical kinetic pathway at high
ATP concentrations is depicted by solid and hollow gray lines.

\begin{figure}
\centerline{\includegraphics[width=3.5in]{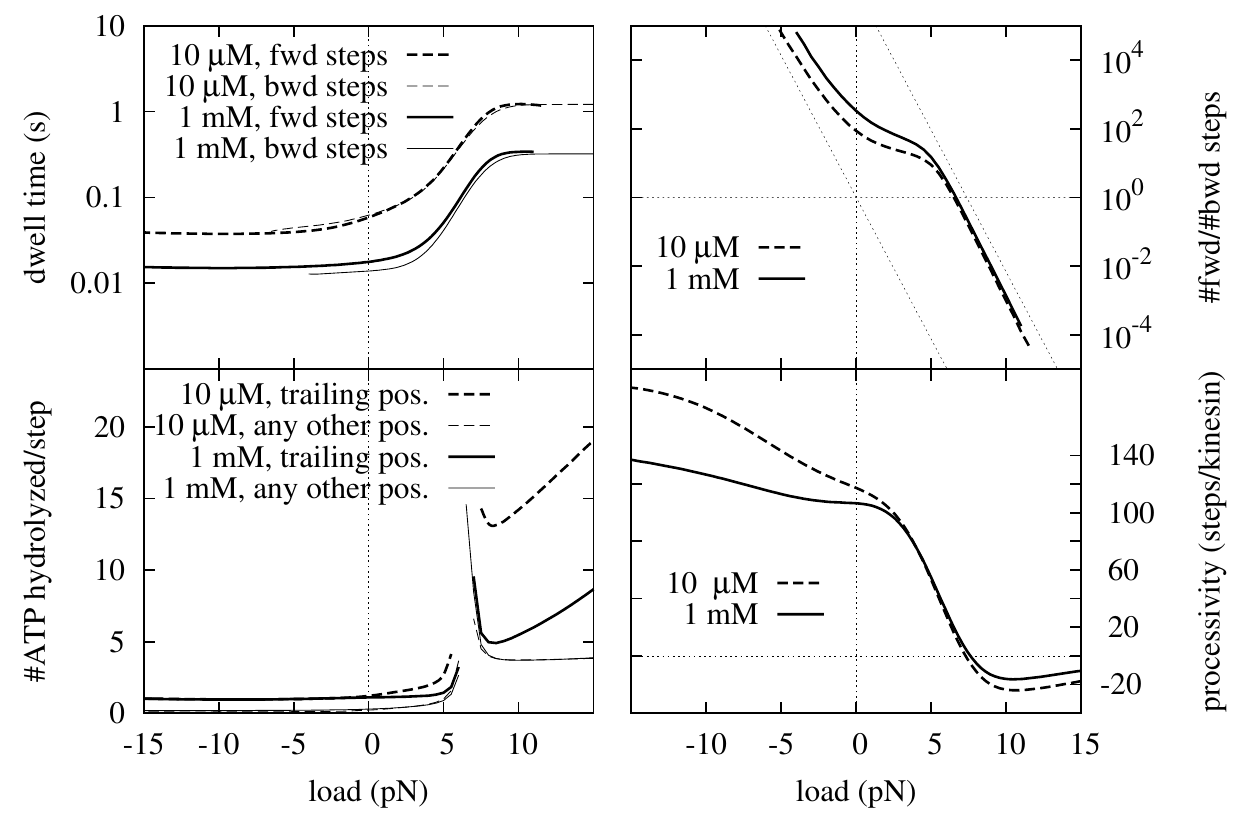}}
\caption{}
\label{fig:carter}
\end{figure}

\subsubsection*{Figure~\ref{fig:carter}.}
Simulation results I:
Several observables at saturating ($1$~mM) and low ($10$~$\mu$M) ATP
concentrations under the full range of external load between $-15$ and
$15$~pN. (Negative load corresponds to assisting force.)

\begin{figure}
\centerline{\includegraphics[width=3.5in]{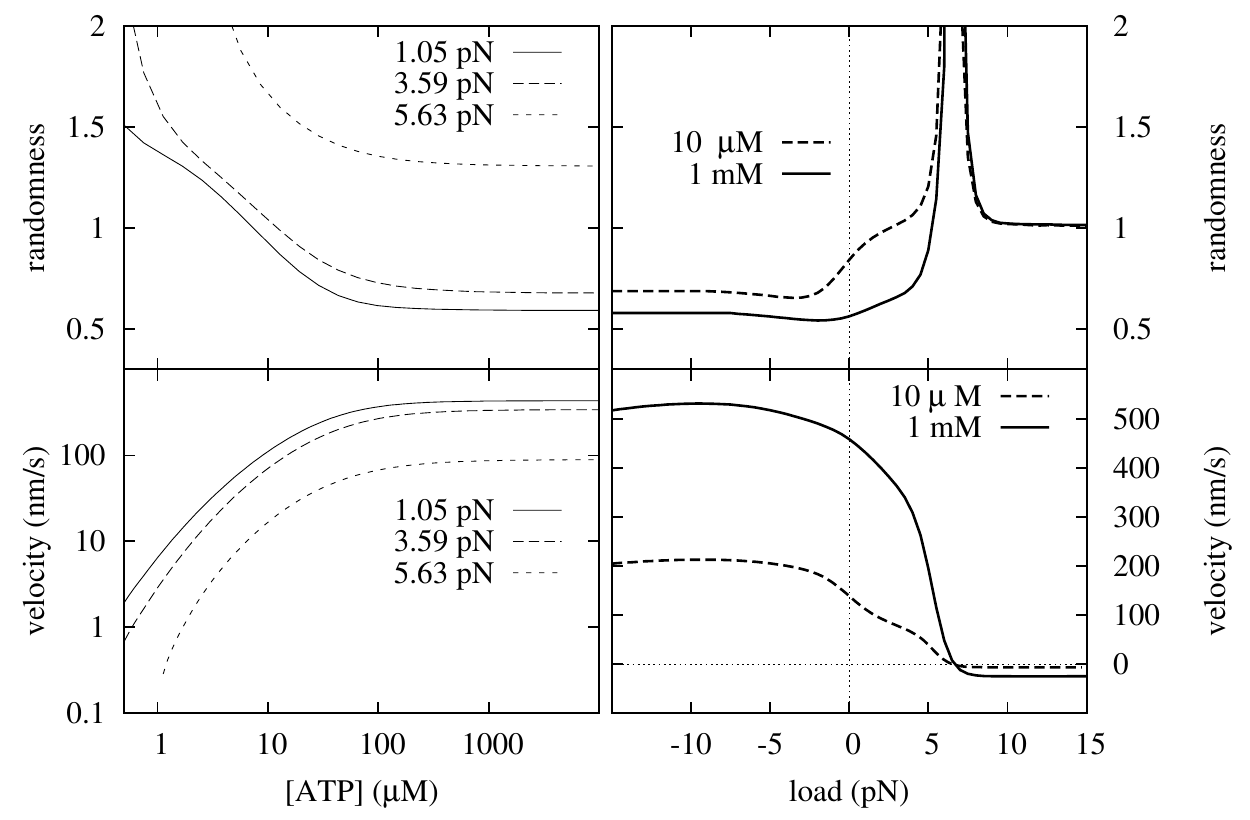}}
\caption{}
\label{fig:ATP_vel_rand}
\end{figure}

\subsubsection*{Figure~\ref{fig:ATP_vel_rand}.}
Simulation results II:
Randomness and velocity plots for two ATP concentrations ($1$~mM and
$10$~$\mu$M) concentrations as functions of the load, and for three
loads ($1.05$, $3.59$, $5.63$ pN) as functions of the ATP concentration.

\begin{figure}
\centerline{\includegraphics[width=3.5in]{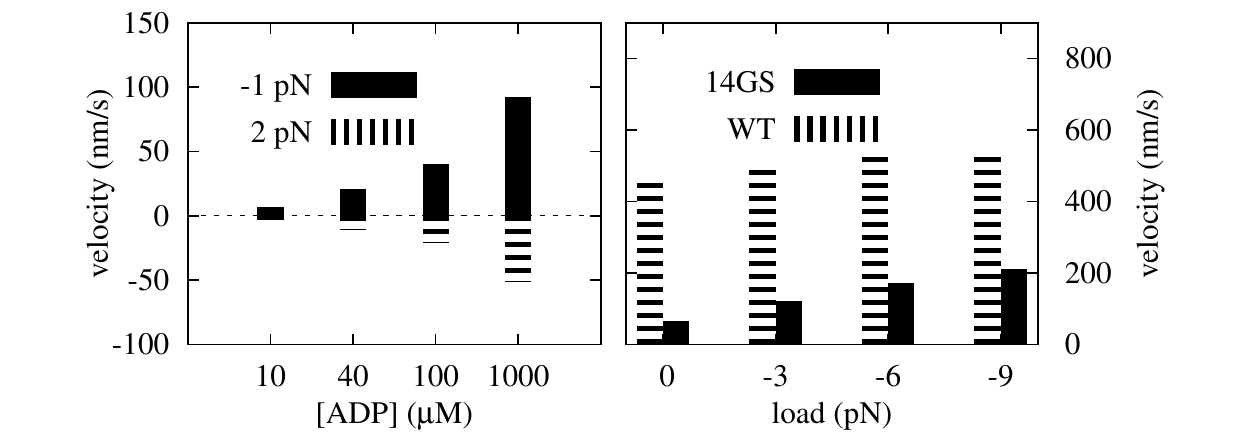}}
\caption{}
\label{fig:michio_14GS}
\end{figure}

\subsubsection*{Figure~\ref{fig:michio_14GS}.}
Simulation results III:
Velocity of kinesin under zero ATP concentration for several
ADP concentrations and external loads (left panel); and the
velocities of the wild type (WT) and the neck linker elongated (14GS)
kinesin for $1$~mM ATP and several external loads (right panel).
%

\clearpage

\clearpage

\clearpage

\clearpage

\clearpage

\end{document}